# ChatGPT Assisting Diagnosis of Neuro-ophthalmology Diseases Based on Case Reports

Yeganeh Madadi, PhD[1], Mohammad Delsoz, MD[1], Priscilla A. Lao, MD[1], Joseph W. Fong, MD[1], TJ Hollingsworth, PhD[1], Malik Y. Kahook, MD[2], and Siamak Yousefi, PhD[1,3]

[1]Department of Ophthalmology, University of Tennessee Health Science Center, Memphis, TN, USA

[2]Department of Ophthalmology, University of Colorado School of Medicine, Aurora, CO, USA

[3]Department of Genetics, Genomics, and Informatics, University of Tennessee Health Science Center, Memphis, TN, USA

Correspondence to: Siamak Yousefi, 930 Madison Ave., Suite 471, Memphis, TN 38163, USA, Phone: 9014487831, Email: Siamak.Yousefi@uthsc.edu

Corresponding author hereby declare that this manuscript is not under simultaneous consideration for publication in another journal nor has it been published elsewhere.

Acknowledgement: This work was supported by NIH Grants R01EY033005 (SY), R21EY031725 (SY), and grants from Research to Prevent Blindness (RPB), New York (SY). The funders had no role in study design, data collection and analysis, decision to publish, or preparation of the manuscript.

Conflicts of interest/Competing interests: Authors declare no relevant conflict of interest(s) to disclose.

**Keywords:** ChatGPT, Large Language Models (LLM), Artificial Intelligence (AI), Neuro-Ophthalmology, Diagnosis

# ABSTRACT

**Purpose:** To evaluate the efficiency of large language models (LLMs) including ChatGPT to assist in diagnosing neuro-ophthalmic diseases based on case reports.

**Design:** Prospective study

**Subjects or Participants:** We selected 22 different case reports of neuro-ophthalmic diseases from a publicly available online database. These cases included a wide range of chronic and acute diseases that are commonly seen by neuro-ophthalmic sub-specialists.

**Methods**: We inserted the text from each case as a new prompt into both ChatGPT v3.5 and ChatGPT Plus v4.0 and asked for the most probable diagnosis. We then presented the exact information to two neuro-ophthalmologists and recorded their diagnoses followed by comparison to responses from both versions of ChatGPT.

**Main Outcome Measures:** Diagnostic accuracy in terms of number of correctly diagnosed cases among diagnoses.

**Results**: ChatGPT v3.5, ChatGPT Plus v4.0, and the two neuro-ophthalmologists were correct in 13 (59%), 18 (82%), 19 (86%), and 19 (86%) out of 22 cases, respectively. The agreement between the various diagnostic sources were as follows: ChatGPT v3.5 and ChatGPT Plus v4.0, 13 (59%); ChatGPT v3.5 and the first neuro-ophthalmologist, 12 (55%); ChatGPT v3.5 and the second neuro-ophthalmologist, 12 (55%); ChatGPT Plus v4.0 and the first neuro-ophthalmologist, 17 (77%); ChatGPT Plus v4.0 and the second neuro-ophthalmologist, 16 (73%); and first and second neuro-ophthalmologists 17 (17%).

**Conclusions**: The accuracy of ChatGPT v3.5 and ChatGPT Plus v4.0 in diagnosing patients with neuro-ophthalmic diseases was 59% and 82%, respectively. With further development, ChatGPT Plus v4.0 may have potential to be used in clinical care settings to assist clinicians in providing quick, accurate diagnoses of patients in neuro-ophthalmology. The applicability of using LLMs like ChatGPT in clinical settings that lack access to subspeciality trained neuro-ophthalmologists deserves further research.

**Summary Highlights:**
- The goal of this study was to explore the capabilities of ChatGPT for the diagnoses of different neuro-ophthalmic diseases using specific case examples.
- There was general agreement between ChatGPT Plus v4.0 and two neuro-ophthalmologists in final diagnoses.
- ChatGPT was more general while neuro-ophthalmologists were more methodical and specific when listing diagnoses.

**INTRODUCTION**

Neuro-ophthalmology is a subspecialty that bridges the fields of neurology and ophthalmology.[1] This specialized discipline focuses primarily on conditions affecting visual pathways, visual processing, and eye movements. Practitioners in this subspecialty have completed residency in neurology or ophthalmology and then an additional year of fellowship training in neuro-ophthalmology. Neuro-ophthalmology is a cognitively intense field, requiring a detailed history and examination and synthesis of neuro-imaging results and lab work, to diagnose often vision- and life-threatening conditions. Neuro-ophthalmologists often are a "specialist's specialist," where many patients are referred by ophthalmologists or neurologists. In a prospective study, nearly half of all referrals to neuro-ophthalmology carried an incorrect diagnosis, and a quarter of these misdiagnosed patients suffered harm.[2]

To add further insult to injury, neuro-ophthalmologists are in short supply, deemed a "human resource crisis" since at least 2008.[3] As of 2022, only 8 states in the United States have adequate coverage and the median wait time to see a neuro-ophthalmologist is 6 weeks.[4]

In recent years, artificial intelligence (AI) and machine learning (ML) technologies have provided promising solutions in various healthcare domains.[5] Several studies have shown the efficacy of AI models in ophthalmology.[6-8] More recently, emerging large language models (LLMs) including ChatGPT (OpenAI, LLC, California, USA), have garnered a great deal of attention for their ability to comprehend and generate human-like and fluent text responses to queries.[9] The commercially available ChatGPT Plus 4.0 has set a new standard for text generation with unparalleled fluency and human-likeness that stretches these limits even further than its predecessor, ChatGPT 3.5. It is worth mentioning that ChatGPT 3.5 was initially trained based on an impressive corpus of over 400 billion words gathered from the web, including books, articles, and diverse digital content.

ChatGPT (accessible at https://chat.openai.com), leverages the architecture of the ChatGPT Plus 4.0 and responds to textual input with high precision and nuance due to improved fine-tuning. The ChatGPT iteration of ChatGPT Plus 4.0 is endowed with an expanded toolkit of natural language processing (NLP) features, ranging from more accurate translation to sophisticated text summarization, demonstrating OpenAI's commitment to advancing the state of conversational AI.

Efforts now have been made to leverage ChatGPT's capabilities in specialized domains, such as medical diagnostics, in addition to general text interactions.[10] However, there are handful research studies discussing applications of ChatGPT in Ophthalmology[11, 12, 13, 14] and the efficacy and reliability of ChatGPT in the specialized field of ophthalmology including neuro-ophthalmology have not been adequately investigated. This paper aims to evaluate ChatGPT's diagnostic capabilities in the field of neuro-ophthalmology. We intend to investigate the model's capabilities and limitations by comparing responses to detailed case descriptions from several patients with various neuro-ophthalmic conditions.

**METHODS**

**Case Collection**

We utilized cases from the Department of Ophthalmology and Visual Sciences at the University of Iowa's publicly accessible database (https://webeye.ophth.uiowa.edu/eyeforum/cases.html). We selected 22 cases with various common and uncommon neuro-ophthalmic diseases including "Acute Demyelinating Encephalomyelitis (ADEM) with Associated Optic Neuritis", "Optic Nerve Drusen", "Optic Nerve Hypoplasia", "Optic Neuritis", "Optic Nerve Hypoplasia", "Cranial Nerve IV (Trochlear Nerve) Palsy", "Dorsal Midbrain Syndrome (Parinaud's Syndrome)", "Traumatic Optic Neuropathy", "Ethambutol Toxicity and Optic Neuropathy", "Posterior Ischemic Optic Neuropathy", "Recurrent Neuroretinitis", "Horner's Syndrome due to Cluster Headache", "Horner Syndrome due to Ipsilateral Internal Carotid Artery Dissection", "Idiopathic Intracranial Hypertension (Pseudotumor Cerebri)", "Idiopathic Orbital Myositis", "Leber Hereditary Optic Neuropathy", "Miller Fisher Syndrome", "Myasthenia Gravis", "Neurofibromatosis Type 1—Optic Nerve Glioma", "Thyroid Eye Disease (Graves' ophthalmopathy)", "Visual Snow Syndrome", and "Unilateral Optic Nerve Hypoplasia". Descriptions of each case included patient demographics, history of the presenting illness, chief complaint, relevant medical or ocular history, and examination findings.

Our local institutional review board (IRB) office advised that no IRB approval is required for this study because we utilized a publicly available dataset containing no patient information and the study adhered to the Helsinki declaration's principles.

**ChatGPT for Diagnosis**

We feed each case description into the ChatGPT (versions 3.5 and 4.0) and asked if the model could provide a diagnosis. We specifically asked: "What is the most probable diagnosis?" (Figures 1 and 2). Then we evaluated the accuracy of ChatGPTs in making a correct diagnosis. As ChatGPT may learn from previous interactions once a single chatBox is used, we recorded all responses based on our first inquiry of diagnosis and used a new chatBox in any case that we asked same questions more than once.

Additionally, we provided the same 22 cases to two neuro-ophthalmologists in our institute in a masked manner and asked them to make provisional diagnosis. Neuro-ophthalmologists were not allowed to use ChatGPT or other similar tools for making diagnosis. We then computed the frequency of correct diagnoses of ChatGPTs and the two neuro-ophthalmologists and compared the accuracies. We also evaluated the agreement between ChatGPTs and the two neuro-ophthalmologists.

**RESULTS**

ChatGPT 3.5 and ChatGPT Plus 4.0 made the correct diagnosis in 13 (59%) and 18 (82%) out of 22 cases, respectively, while two neuro-ophthalmologists both correctly diagnosed 19 (86%) out of 22 cases. Table 1 represents the details of diagnoses which are provided by ChatGPT 3.5, ChatGPT Plus 4.0 and the two neuro-ophthalmologists.

The agreement between the various diagnostic sources were as follows: ChatGPT v3.5 and ChatGPT Plus v4.0, 13 (59%); ChatGPT v3.5 and the first neuro-ophthalmologist, 12 (55%); ChatGPT v3.5 and the second neuro-ophthalmologist, 12 (55%); ChatGPT Plus v4.0 and the first neuro-ophthalmologist, 17 (77%); ChatGPT Plus v4.0 and the second neuro-ophthalmologist, 16 (73%); and first and second neuro-ophthalmologists 17 (17%).

**DISCUSSION**

We prospectively investigated the diagnostic capability of ChatGPT in neuro-ophthalmic diseases based on 22 cases collected from the University of Iowa online database. ChatGPT 3.5, ChatGPT Plus 4.0, and both neuro-ophthalmologists diagnosed 13, 18, 19, and 19 cases correctly out of 22 cases, respectively. Despite the complexity of neuro-ophthalmic cases, the diagnostic capability of ChatGPT Plus 4.0 was close to neuro-ophthalmologists. Under active training, it is expected that newer versions of ChatGPT may become more accurate in the future.

While the agreements between ChatGPT v3.5 and neuro-ophthalmologists were relatively low (55% - 59%), ChatGPT Plus v4.0 had a reasonable agreement with neuro-ophthalmologists (73% - 77%). The agreement between the neuro-ophthalmologists was 77%, on par with ChatGPT Plus 4.0.

Understanding capabilities and limitations of ChatGPT will help to effectively utilize this technology in neuro-ophthalmology research and clinical practice.

A recent study assessed the performance of ChatGPT in answering practice questions in neuro-ophthalmology and observed that ChatGPT was correct on 3 out of 7 responses (43%) of the text-based questions.[15] Another study investigated ChatGPT 3.5 and ChatGPT Plus 4.0 capabilities in responding to various ophthalmological questions including neuro-ophthalmic questions based on the online database of OphthoQuestions and the American Academy of Ophthalmology's Basic Clinical Science Course (BCSC). They identified that the accuracy of ChatGPT 3.5 was 10% and 32% and the accuracy of ChatGPT Plus 4.0 was 40% and 48% based on the OphthoQuestions and BCSC neuro-ophthalmic questions, respectively.[11] Their reported accuracy in ChatGPT capability in responding to neuro-ophthalmic questions is significantly lower than what we observed in neuro-ophthalmic diagnosis, however, the nature of these two studies are different, making it challenging to directly comparing the findings of these two studies.

Advancement of LMM models including ChatGPT may benefit ophthalmology in general and neuro-ophthalmology in particular by making more objective and quick diagnosis based on readily accessible models. Neuro-ophthalmic issues result from a complex interaction between the nervous system and visual apparatus, necessitating a comprehensive evaluation to identify subtle distinctions.[16]

Another significant advantage of ChatGPT and comparable LMM models is that they can actively learn based on their reinforcement learning capability, thereby correcting their previous errors, enhancing their performance over time, and becoming more precise. In addition, LMM models require minimal human monitoring and supervision for active training, which is advantageous in comparison to supervised learning models that require intensive human oversight.

While we used patients with various phenotypes of neuro-ophthalmic conditions, our study does have some limitations. First, we used cases from an online database to investigate ChatGPT capabilities raising concerns that ChatGPT may have previously been exposed to this database. To resolve this issue, we examined the years in which these cases were added to the database and found that the vast majority were added before September 2021, when the most recent ChatGPT training concluded. Versions 3.5 and 4.0 of ChatGPTs contained errors in 9 and 4 instances, respectively, alleviating concerns that this database has been accessed previously. Second, ChatGPT has been evaluated based on 22 cases. Consequently, other studies may evaluate ChatGPT on the basis of a larger number of cases to confirm our findings. Another limitation of LLMs like ChatGPT is lack of image interpretation that limits their applicability in more complex cases with information from various imaging modalities. Additionally, while these were real world cases, they were published primarily for teaching purposes. Several of these cases exhibited classic presentations of diseases with "buzzword" descriptions with definitive diagnoses. Therefore, this subset of cases may not fully reflect the actuality of neuro-ophthalmology practice with borderline findings and confounding factors in many instances. Conversely, some cases exhibited atypical presentations of the actual diagnosis. This is where the art of medicine comes into play and requires a human practitioner to weigh these factors appropriately.

In summary, various AI models have shown great potential in ophthalmology. Emerging ChatGPT has been recently applied to ophthalmic education as well as diagnosis and has been shown to be capable of conversing with patients as well as physicians and respond to text inquiries with fluent and reasonable accuracy. These models may also be utilized in a variety of healthcare settings, including primary care offices or emergency services to aid in patient triaging, with the caveat that not all patient information is available in such settings. These models can also be useful in tertiary ophthalmology care, to provide initial assessments. Once capable of interpreting images in the future, LMMs including ChatGPT may provide more objective, rapid, and possibly accurate diagnosis based on both textual and imaging input.

**CONCLUSION**

The introduction of LMMs, such as ChatGPT, represents a potentially revolutionary step forward augmenting clinical diagnostic, independent of the availability of subspeciality trained clinicians. This is especially interesting in fields such as neuro-ophthalmology, which is characterized by an abundance of complicated cases and the absence of a conclusive diagnostic testing. ChatGPT Plus v4.0 appears capable of diagnosing complex neuro-ophthalmic cases, when provided with structured data in case report format, with a level of accuracy that is comparable to that of experienced neuro-ophthalmologists. Re-assessment of the performance of ChatGPT, and other LLMs, in similar testing conditions but with multimodal capabilities would be a natural next research step when practical and pending availability of such algorithms. LLMs could potentially bridge the gap between AI and human competence. Future research should explore how technological innovations, such as enhanced LLMs, can augment medical decision making and assist clinicians towards enhancing patient outcomes.

**Figures and Tables**

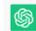

Figure 1. A sample case description of a patient with idiopathic intracranial hypertension (pseudotumor cerebri) input into the ChatGPT 3.5 model and the corresponding response.

[ChatGPT conversation screenshot]

**User (GPT-4):** A 31-year-old obese female presented with complaints of headaches and transient visual obscurations that had been occurring for six months, along with a ringing sound in her ears that was synchronous with her pulse, persisting for a year. On examination, she was found to have bilateral optic disc edema. An MRI scan was performed, revealing an empty sella but no other abnormalities; an MR venogram was normal with no evidence of sinus thrombosis. A lumbar puncture indicated an elevated opening pressure of 50 cm H2O, leading to a prescription of a carbonic anhydrase inhibitor at 500 mg twice a day, along with advice to lose weight. However, the patient continued to experience severe headaches, and one month later, visual field testing displayed significant mean deviations in both eyes along with enlargement of the blind spot. A repeat lumbar puncture confirmed the elevated pressure, resulting in an increased medication dosage to 500 mg three times daily. The patient's past medical history includes obesity and depression, with a daily intake of 1500 mg of the prescribed medication and citalopram hydrobromide. There's no significant family history, and her social history includes non-smoking and no alcohol use, although she reported substantial weight fluctuations in recent months. Her physical exam revealed visual acuity of OD-- 20/20 and OS-- 20/50, full extraocular motility without diplopia, pupils at 5 mm in the dark and 2 mm in the light with no relative afferent pupillary defect, intra-ocular pressure of 11 mmHg OD and 16 mmHg OS, normal color vision, normal cranial nerve V and VII function bilaterally, a normal external and slit lamp exam, and grade two papilledema with otherwise normal ocular findings. What is the most probable diagnosis?

**ChatGPT:** I am not a medical professional, but the presentation of headaches, transient visual obscurations, pulsatile tinnitus, and elevated opening pressure on lumbar puncture along with the other findings such as bilateral optic disc edema and an empty sella on MRI are highly suggestive of Idiopathic Intracranial Hypertension (IIH), also known as Pseudotumor Cerebri. This condition is characterized by elevated intracranial pressure with no evidence of a tumor or other diseases affecting the brain. It is most commonly found in obese women of childbearing age, and the risk factors include obesity and female gender, both of which are present in this case.

A normal MR venogram that shows no evidence of sinus thrombosis further supports the diagnosis by ruling out secondary causes of raised intracranial pressure. In this case, the lumbar puncture with elevated opening pressure further confirms the diagnosis.

Given the persistence and worsening of symptoms (headache and worsening visual field test results) despite medical treatment (carbonic anhydrase inhibitor), it would be advisable to consult a medical team comprising neurologists, neuro-ophthalmologists, and possibly neurosurgeons for an urgent review and possible surgical interventions such as optic nerve sheath fenestration or a cerebrospinal fluid shunting procedure, to prevent permanent vision loss.

The presence of grade two papilledema is a concerning sign, especially in combination with the worsening visual field tests, suggesting a risk of permanent vision loss if not properly managed. Close monitoring and treatment are essential.

Figure 2. A sample case description of a patient with idiopathic intracranial hypertension (pseudotumor cerebri)) input into the ChatGPT Plus 4.0 model and the corresponding response.

Table 1. Diagnoses provided by ChatGPT 3.5, ChatGPT Plus 4.0, and two neuro-ophthalmologists.

| Name of disease | ChatGPT 3.5 diagnosis | ChatGPT Plus 4.0 diagnosis | Neuro-ophthalmologist 1 diagnosis | Neuro-ophthalmologist 2 diagnosis |
|---|---|---|---|---|
| Demyelinating Encephalomyelitis (ADEM) with Associated Optic Neuritis | Disseminated Encephalomyelitis (ADEM) | Neuromyelitis Optica Spectrum Disorder (NMOSD) | Optic Neuritis 2/2 Meningitis | Viral Meningitis |
| Optic Nerve Drusen | Papilledema, specifically due to Idiopathic Intracranial Hypertension (IIH) | Optic Disc Drusen | Optic Disc Drusen | Optic Disc Drusen |
| Optic Nerve Hypoplasia | Congenital Glaucoma | Mature Cataract and Optic Atrophy in the Left Eye, along with Exotropia | Optic Nerve Hypoplasia | Optic Nerve Hypoplasia |
| Optic Neuritis | Optic Neuritis | Optic Neuritis | Optic Neuritis 2/2 MS | Multiple Sclerosis, Optic Neuritis |
| Optic Nerve Hypoplasia | Septo-Optic Dysplasia (SOD) | Optic Nerve Hypoplasia (ONH) | Demorsier Syndrome | Septo-Optic Dysplasia with Optic Nerve Hypoplasia |
| Cranial Nerve IV (Trochlear Nerve) Palsy | Fourth Nerve Palsy | Fourth Nerve Palsy (Trochlear Nerve Palsy) | Traumatic CN IV Palsy OD | Right Traumatic Fourth Nerve Palsy |
| Dorsal Midbrain Syndrome (Parinaud's Syndrome) | Superior Oblique Myokymia | Parinaud's Syndrome, (Dorsal Midbrain Syndrome) | Dorsal Midbrain Syndrome | Dorsal Midbrain Syndrome |
| Traumatic Optic Neuropathy | traumatic optic neuropathy (TON) | Traumatic Optic Neuropathy (TON) | Traumatic Optic Neuropathy 2/2 Orbital Canal Fracture | Traumatic Optic Neuropathy |
| Ethambutol Toxicity and Optic Neuropathy | Ethambutol-Induced Optic Neuropathy | Toxic Optic Neuropathy, most likely related to the use of Ethambutol | Ethambutol Toxicity/Toxic Optic Neuropathy | Toxic (Ethambutol) Optic Neuropathy |
| Posterior Ischemic Optic Neuropathy | Posterior Ischemic Optic Neuropathy (PION) | Perioperative Ischemic Optic Neuropathy (PION) | PION | Posterior Ischemic Optic Neuropathy, Bilateral |
| Recurrent Neuroretintitis | Multiple Sclerosis (MS) | Optic Neuritis (ON) | Optic Neuritis | Optic Neuritis |
| Horner's Syndrome due to Cluster Headache | Cluster Headache, specifically the variant known as "Cluster-Tic Syndrome" or "Cluster Migraine Syndrome | Cluster Headache (CH), specifically with a Horner's syndrome | Horner Syndrome 2/2 Cluster Headaches | Cluster Headache |
| Horner Syndrome due to Ipsilateral Internal Carotid Artery Dissection | Carotid Artery Dissection | Horner's Syndrome related to a Carotid Artery Dissection | Horner Syndrome 2/2 Carotid Dissection | Cervical Carotid Artery Dissection Causing Horner Syndrome |
| Idiopathic Intracranial Hypertension (Pseudotumor Cerebri) | Idiopathic Intracranial Hypertension (IIH), also known as Pseudotumor Cerebri | Idiopathic Intracranial Hypertension (IIH), also known as Pseudotumor Cerebri | IIH | Idiopathic Intracranial Hypertension |
| Idiopathic Orbital Myositis | Graves' ophthalmopathy, also known as thyroid eye disease | Idiopathic Orbital Inflammatory Syndrome (IOIS), also known as Orbital Myositis | Orbital Myositis | Thyroid Eye Disease |
| Leber Hereditary Optic Neuropathy | Idiopathic Intracranial Hypertension (IIH) | Papilledema | LHON | Leber's Hereditary Optic Neuropathy |
| Miller Fisher Syndrome | Miller Fisher Syndrome | Miller Fisher Syndrome (MFS) | Miller Fisher/Guillan Barre | Miller Fisher Syndrome |
| Myasthenia Gravis | Myasthenia Gravis (MG) | Myasthenia Gravis (MG) | Myasthenia Gravis | Ocular Myasthenic Gravis |
| Neurofibromatosis Type 1—Optic Nerve Glioma | Optic Pathway Glioma | Neurofibromatosis Type 1 (NF1) | Optic Nerve Glioma | Optic Pathway Glioma |
| Thyroid Eye Disease (Graves' Ophthalmopathy) | Thyroid Eye Disease (Graves' Ophthalmopathy) | Thyroid Eye Disease (TED), also known as Graves' Ophthalmopathy | TED | Thyroid Eye Disease with Compressive Optic Neuropathy |
| Visual Snow Syndrome | Persistent Migraine Aura Without Infarction (PMA) | Visual Snow Syndrome (VSS) | Visual Snow | Visual Snow Syndrome |
| Unilateral Optic Nerve Hypoplasia | Optic Nerve Hypoplasia (ONH) | Optic Nerve Hypoplasia (ONH) | Sensory Exotropia 2/2 Optic Nerve Hypoplasia | Optic Nerve Hypoplasia |